\documentclass[showpacs,twocolumn,aps,superscriptaddress,preprintnumbers,letterpaper]{revtex4}
\usepackage{amsmath,amssymb}
\usepackage{epsfig}
\usepackage{graphicx}
\usepackage{amsmath}
\usepackage{amsfonts}
\usepackage{epstopdf}

\newcommand{\be}{\begin{equation}}
\newcommand{\ee}{\end{equation}}
\newcommand{\bear}{\begin{eqnarray}}
\newcommand{\eear}{\end{eqnarray}}
\newcommand{\ba}{\begin{array}}
\newcommand{\ea}{\end{array}}

\def\be{\begin{eqnarray}}
\def\ee{\end{eqnarray}}

\def\roughly#1{\mathrel{\raise.3ex\hbox{$#1$\kern-.75em%
\lower1ex\hbox{$\sim$}}}}

\def\la{{\Big<}}
\def\ra{{\Big>}}

\def\J#1#2#3#4{ {#1} {\bf #2} (#4) {#3}. }

\def\PLB{Phys. Lett. B}

\begin{document}

\title{Holographic Pomeron and the Schwinger Mechanism}

\author{G\"ok\c ce Ba\c sar}
\email{basar@tonic.physics.sunysb.edu}
\affiliation{Department of Physics and Astronomy, Stony Brook University, Stony Brook, New York 11794-3800, USA}

\author{Dmitri E. Kharzeev}
\email{Dmitri.Kharzeev@stonybrook.edu}
\affiliation{Department of Physics and Astronomy, Stony Brook University, Stony Brook, New York 11794-3800, USA}
\affiliation{Department of Physics, Brookhaven National Laboratory, Upton, New York 11973-5000, USA}

\author{Ho-Ung Yee}
\email{hyee@tonic.physics.sunysb.edu}
\affiliation{Department of Physics and Astronomy, Stony Brook University, Stony Brook, New York 11794-3800, USA}

\author{Ismail Zahed}
\email{zahed@tonic.physics.sunysb.edu}
\affiliation{Department of Physics and Astronomy, Stony Brook University, Stony Brook, New York 11794-3800, USA}

\date{\today}

\begin{abstract}

We revisit the problem of dipole-dipole scattering via exchanges of soft Pomerons in the context of holographic QCD.
We show that a single closed string exchange contribution to the eikonalized dipole-dipole
scattering amplitude yields a Regge behavior of the elastic amplitude; the corresponding slope and intercept are different from previous
results obtained by a variational analysis of semi-classical surfaces. 
We provide a physical interpretation of the semi-classical worldsheets driving the Regge behavior for $(-t)>0$ in terms of
worldsheet instantons. The latter describe the Schwinger mechanism for string pair creation by an electric field, where
the longitudinal electric field $E_L=\sigma_T\,{\rm tanh}(\chi/2)$ at the origin of this
non-perturbative mechanism is induced  by the relative rapidity $\chi$ of the scattering dipoles.
Our analysis naturally explains the diffusion in the impact parameter space encoded in the Pomeron exchange; in our picture, it is due to the Unruh temperature of 
accelerated strings under the electric field. We also argue for the existence of  a "micro-fireball'' in the middle of the transverse space
due to the soft Pomeron exchange, which may be at  the origin of the thermal character of multiparticle production in ep/pp collisions.  
After summing over uncorrelated multi-Pomeron exchanges, we find that   
the total dipole-dipole cross section obeys the Froissart unitarity bound.
\end{abstract}
\pacs{12.39.Fe,12.38.Qk,13.40.Em}

\maketitle

\setcounter{footnote}{0}

\section{Introduction}

Near-forward parton-parton and dipole-dipole scattering 
at high energies is sensitive to the infared
aspects of QCD. 
General QCD arguments show that the 
resummation of a class of t-channel exchange gluons
may account for the reggeized form of the scattering
amplitude~\cite{BFKL}, qualitatively consistent with the observed
 growth of the scattering amplitude~\cite{EXPERIMENT}. Nevertheless some empirical 
 features of the hadron-hadron scattering (e.g. the Pomeron slope) point to the importance of non-perturbative effects.

A nonperturbative formulation of high-energy scattering
in QCD was originally suggested by Nachtmann~\cite{NACHTMANN} 
and others~\cite{VERLINDE,KORCHEMSKY} using arguments in 
Minkowski space. At high energy, the near-forward scattering
amplitude can be reduced to a correlation function of two
Wilson lines (parton-parton) or Wilson loops (dipole-dipole)
in the QCD vacuum. The pertinent correlation was assessed
in leading order using a two-dimensional sigma model with
conformal symmetry~\cite{VERLINDE}, and also the anomalous
dimension of the cross-singularity between the two Wilson
lines~\cite{KORCHEMSKY}. Both analyses were carried out in 
Minkowski geometry, with a close relation to QCD perturbation
theory.

An Euclidean formulation was used within the stochastic vacuum
model through a cumulant expansion in~\cite{DOSCH} 
to assess the Wilson loop correlators in Euclidean space. 
Their phenomenological relevance to proton-proton scatering
was pursued in~\cite{PIRNER}.
The instanton vacuum approach to the parton-parton and dipole-dipole
scattering amplitudes was used in~\cite{SHURYAK} to estimate the role
of instanton-antinstanton configurations in both the elastic and inelastic 
amplitudes. In particular, a class of singular gauge configurations reminiscent of 
QCD sphalerons were shown to be at the origin of the inelasticities. The smallness 
of the pomeron intercept was shown to follow from the smallness of the instanton
packing fraction in the QCD vacuum.
The ``instanton ladder"  has been argued to generate the soft Pomeron both at weak 
~\cite{Kharzeev:1999vh,KHARZEEV} and strong coupling, through D-instantons \cite{Kharzeev:2009pa}. 
 First principle considerations of the 
Wilson-line correlators in Euclidean lattice gauge theory have now
appeared in~\cite{MEGGIOLARO} which may support the arguments
for non-perturbative physics in diffractive processes.

Elastic and inelastic scattering in holography have been addressed initially 
in the context of the conformally symmetric AdS$_5$ setting using Minkowskian string surface 
exchanges between the Wilson-line/loops  in the eikonal approximation~\cite{Rho:1999jm}. 
This approach was further exploited in~\cite{Janik:2000aj,Janik:2000pp,Janik:2001sc,Giordano:2011sn} to address 
the same problem in holographic QCD with confinement~\cite{WITTEN} for quark-antiquark scattering.  
In the confined Euclidean background geometry, it was assumed that the most part of string worldsheet stays at the infrared (IR) end point
in the holographic direction, so that the problem effectively reduces to the flat space one with an effective string tension at the IR end point. The helicoidal surface was argued as 
the minimal string surface between two Wilson lines for large impact parameter. The 
inelasticities (a deviation of the amplitude from being a pure phase) were identified through a multibranch structure in
analytic continuation from Euclidean to Minkowski space.  However, the physics picture behind this multibranch structure has been somewhat mysterious. 
One-loop string fluctuations around the
helicoidal surface have shown to be important for addressing key aspects of the
Pomeron and Reggeon physics such as intercepts. 

A more thorough study of the Pomeron problem 
in the context of holography has been performed in~\cite{TAN}.  Specifically, the Pomeron was argued to follow from 
a full string amplitude in a curved geometry of holographic QCD, including fluctuations in the holographic radial direction. 
One of our motivations for the present work is to clarify the relation between the approaches in \cite{Rho:1999jm,Janik:2000aj,Janik:2000pp,Janik:2001sc,Giordano:2011sn} and the one in \cite{TAN}, identifying the valid regime of approximations in the analysis of the former.

A compelling picture of the role of the holographic radial direction as one varies $t=-q^2$ was presented in \cite{TAN}. As $(-t)\gg M_{KK}^2$, where $M_{KK}$ denotes a mass scale of confinement, the string worldsheet was shown to be pushed to the UV regime along the holographic direction where the behavior of Pomeron kernel becomes similar to the BFKL~\cite{Lipatov:1976zz,Kuraev:1977fs,Balitsky:1978ic}. The regime $(-t)\le M_{KK}^2$
is however more model-dependent, and the string worldsheet can in principle stay close to the IR end point. 
It is in this regime (soft Pomeron regime) that the flat-space approximation in \cite{Janik:2000aj,Janik:2000pp,Janik:2001sc,Giordano:2011sn}
can be justified. 

Based on the same flat-space approximation for soft Pomerons, we will attempt to compute a full closed string exchange amplitude
between two Wilson loops in dipole-dipole scattering.
The two Wilson loops with large relative rapidity set the relevant asymptotic states in the high energy eikonal formulation, and provide an effective boundary condition for the exchanged closed strings. For a small dipole size $a$, 
this boundary condition will be argued to be similar to the one in the D0 brane scattering problem, which allows us to 
compute, modulo a few subtle differences, all the essential features of the expected Reggeized amplitude in soft Pomeron regime.

The Regge behavior of closed string exchange in flat space has been known for a while from the simplest Virasoro-Shapiro
amplitude of $2\to2$ scattering. Irrespective of the details of the external states, the closed string exchange gives rise to
a universal Pomeron kernel
\be
s^{2+{\alpha'\over 2} t}\,,\label{kernel}
\ee
where the $2$ in the intercept should be replaced by ${D_\perp \over 12}$ for purely bosonic string, where $D_\perp$ is the number of massless bosonic worldsheet fluctuations minus two from ghosts. 
The universality of the above kernel can be understood from the fact that it arises through semi-classical worldsheets
connecting the two high energy string states, whose lengths in the impact parameter space $b$ are of order $b\sim \sqrt{\ln (s)}\gg 1$
in units of $\sqrt{\alpha'}$~\cite{Gross:1987kza}. As $b$ is large, the details of the external states are not relevant in (\ref{kernel}). A similar conclusion will be reached in our case of dipole Wilson loops specifying the external states, as it should. 

In this work, we intend to clarify a number of physical issues of relevance to dipole-dipole scattering
in the diffractive regime based on stringy holography:
\begin{enumerate}
 \item We would like to clarify the results of \cite{Janik:2000aj,Janik:2000pp,Janik:2001sc}, especially the 
ambiguity of the multibranch structure in the variational approach. The universal result in (\ref{kernel})
indicates that only the minimal branch cut is physical, while higher winding contributions are artifacts of the variational method. Also, the slope $\alpha'\over 4$ and the intercept $1+{D_\perp\over 96}$ in their results seem to differ from (\ref{kernel}).

\item
We would like to understand the regime of validity of the approximations used in those works, such as using the 
flat-space approximation and neglecting the massive worldsheet fermions. We also would like to clarify the relation 
to the analysis in~\cite{TAN}. We will elaborate on this in section \ref{valid}.

\item 
We would like to understand the physical origin of the universal semi-classical worldsheets that are responsible for the Pomeron kernel (\ref{kernel}).
The effective D0 brane scattering analogue we have for a small dipole size $a$ turns out to be useful for this. Indeed,   
via worldsheet T-duality, we show that these 
semi-classical worldsheets map to the
stringy analogue of worldline instantons (anti-instantons) 
in the Schwinger mechanism of pair creation under an external electric field, where
the effective electric field in our T-dual picture is induced by the relative rapidity of the original Wilson loops.
This gives us more insight onto the nature of the semi-classical worldsheets. Moreover, we will argue that the Schwinger mechanism 
description  indicates the  existence of a ``micro-fireball'' in the middle of the created string due to the Unruh temperature of an
accelerated string worldsheet, which may explain the observed thermal multiplicity in pp collisions as well as the diffusion behavior implied by the soft Pomeron.

\item
Beyond the universal kernel (\ref{kernel}), we also include the dependence of the full amplitude on the dipole size $a$, 
in a reasonable approximation. This is a question of prefactor multiplying (\ref{kernel}) that depends on the size (virtuality) of the external dipole states. Our result is reminiscent of the  phenomenological dipole parameterization of the cross section of deep-inelastic scattering in terms of a dipole size $a$ and the saturation momentum \cite{GBW}.

\item We also consider the case of dipole Wilson loops of higher representations. We show that this case allows some of the multi-winding contributions with winding number $k\le k_{\rm max}$, where $k_{\rm max}$ depends on the representation. 
When $k$ becomes comparable to $N_c$, the N-ality becomes important and the correct objects exchanged should be $k$-strings
described by D-branes, instead of simple overlapping $k$ number of strings, so that the results for large $k$ should be modified.

\end{enumerate}

In view of computing the connected expectation value of two largely separated Wilson loops, we comment on one aspect of our result.
Indeed, we note that the  semi-classical worldsheets responsible for (\ref{kernel}) exist for large impact parameters.
They are sustained by the rapidity of the two Wilson loops. In the case of zero rapidity, that is, for a pair
of static Wilson loops, it has been known that there is a phase transition at large distance where 
semi-classical worldsheets connecting the two Wilson loops cease to exist~\cite{Gross:1998gk}, and one necessarily
goes to the perturbative supergravity mode exchanges.
Interestingly, this Gross-Ooguri transition is removed in our case by a finite rapidity difference between the two Wilson loops:
there always exist semi-classical worldsheets between the two Wilson loops with rapidity angle. 
In the T-dual picture, this is due to the fact that a finite electric field always 
admits stringy worldsheet instantons for any separation of two end points of the string. 
One can also check, for example in (\ref{POLES}), that these contributions disappear in a static limit $\chi\to 0$,
conforming to the perturbative supergravity exchange regime, thanks to the occurence of an essential singularity.

Although our results are based on general features of holographic models with confinement,
it is nonetheless useful to have a reference model, expecially when we discuss the regime of validity of our approximations.  We will consider the double Wick-rotated non-extremal D4-brane geometry by Witten~\cite{WITTEN}.
This holographic QCD with D4 branes offers a nonperturbative framework for discussing 
Wilson loops in the double limit of large number of colors $N_c$ and 
t' Hooft coupling  $\lambda=g^2N_c$.
The effective string tension at the IR end point is given by $\sigma_T={2\over 27\pi} M_{KK}^2 \lambda$ 
(or $\alpha'=1/2\pi\sigma_T$), although the expression is model-dependent and not essential for our purposes.

The outline of the paper is as follows: in section \ref{DDPT} we set the definitions
for the eikonalized dipole-dipole scattering amplitudes in the impact parameter space representation, and review their analysis
in Euclidean perturbation theory.  In section \ref{QQPT} we compute the string amplitude of t-channel closed string exchange between the two dipole Wilson loops in holographic QCD, based on a few reasonable assumptions. The Wilson-loop correlation function is shown to pick up a real part (corresponding to inelasticity) from the pole contributions generated by the
rapidity twisting of the bosonic zero modes. We then identify these contributions with the semi-classical worldsheet instantons
in the Schwinger mechanism in the T-dual picture, where the electric field is induced by the relative rapidity.   
We also argue that not all contributions from multiple $k>1$ windings are physical due to a difference between real D0 brane
and our Wilson loops, and one necessarily needs to truncate the sum up to $k=k_{\rm max}$ depending on the representation of the Wilson loops: for the fundamental representation, $k_{\rm max}=1$. We discuss a related interesting issue of $N$-ality and $k$-strings in our picture.
In section \ref{ELAS}, we obtain our elastic dipole-dipole scattering amplitude 
from soft Pomeron exchange in the momentum space, and discuss the phenomenology of our results. The parallel between our Pomeron and the empirical soft Pomeron advocated by Donnachie ans Landshoff are detailed in section~\ref{DON}.
In section \ref{valid}, we examine the validity regime of our assumptions taken in the computation.
We then show in section \ref{Froissart} that the total cross section from the eikonal exponentiation of our results obeys the Froissart unitarity bound.
Our conclusions are in section \ref{conclusion}.

\section{Perturbation Theory}\label{DDPT}

We consider an elastic dipole-dipole scattering
\be
D_1\,(p_1) + D_2\,(p_2) \rightarrow D_1\,(k_1) +D_2\,(k_2)\,,
\label{3X}
\ee
with a dipole size $a$, and $s=(p_1+p_2)^2$, $t=(p_1-k_1)^2$, $s+t+u=4m^2$. 
The color and spin of the incoming/outgoing quarks inside the dipoles are traced over.

In Euclidean signature, the kinematics is fixed by noting that
the Lorentz contraction factor translates to
\bear
{\rm cosh}\,\chi  =\frac s{2m^2}-1\rightarrow {\rm cos}\,\theta\,\,,
\label{LO1}
\eear
where $\theta$ is the Euclidean angle between the two high energy trajectories in the longitudinal space and ${\rm cosh}(\chi/2)=\gamma= (1-v^2)^{-1/2}$ in the center of mass frame.
Scattering at high-energy in Minkowski geometry follows from analytically continuing $\theta\rightarrow -i\chi$ 
in the regime $\chi\approx \,{\rm log}\, (s/2m^2)\gg 1$~\cite{THETAMEGGIOLARO}. 
It is convenient to consider the trajectories in the impact space representation as  
$p_1/m=\,({\rm cos}(\theta/2)\,,-{\rm sin}(\theta/2),\,0_\perp)$, $p_2/m=\, 
({\rm cos}(\theta/2)\,,{\rm sin}(\theta/2),\,0_\perp)$, $q=(0,0,q_\perp)$
and $b=(0,0,b_\perp)$, where $q$ is the t-channel momentum ($t=-q_\perp^2<0$) and $b$ is the impact parameter.
(The first two coordinates are longitudinal space and the $\perp$ collectively means the transverse two dimensional impact parameter space).

\begin{figure}
\includegraphics[scale=0.6]{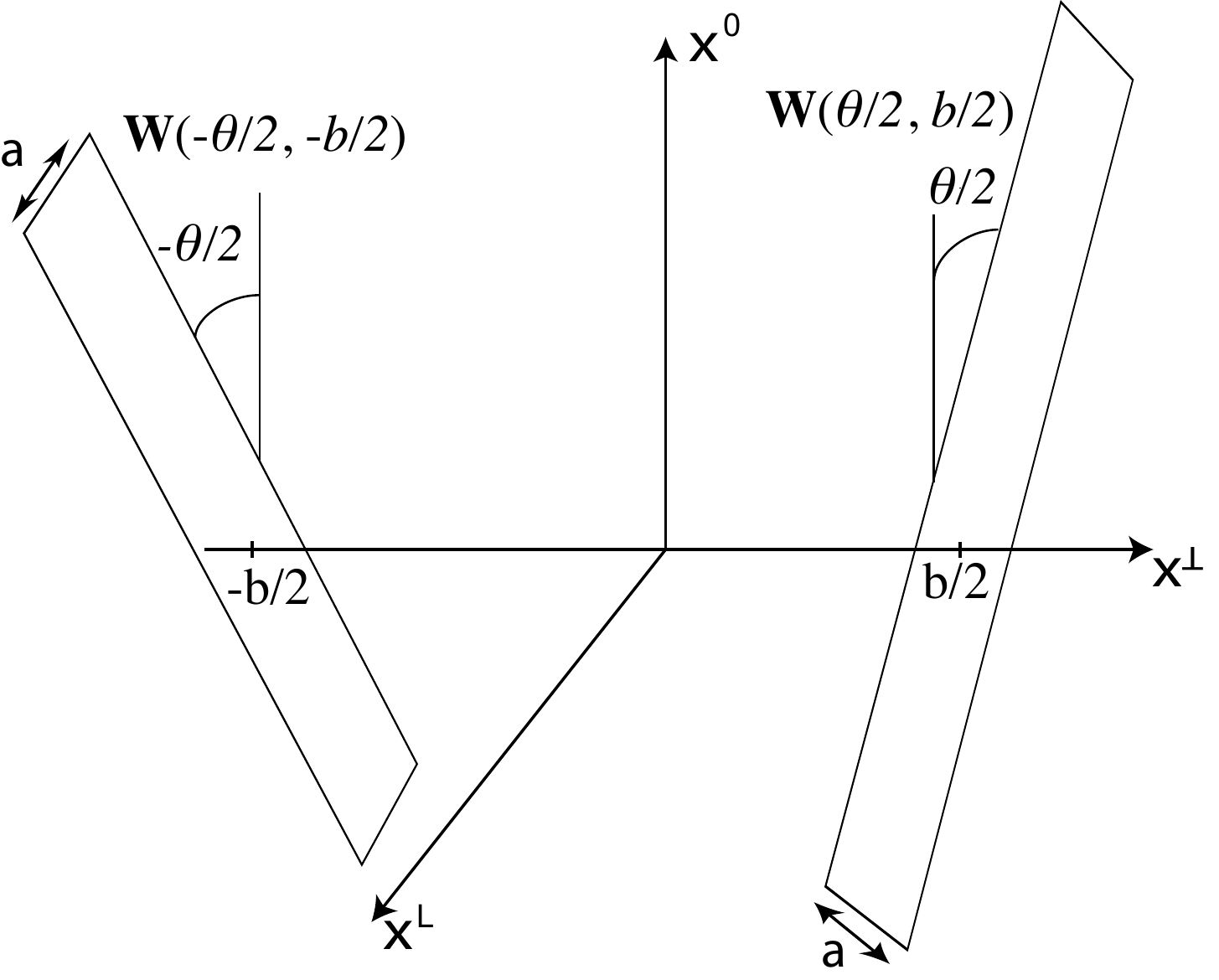}
\caption{Dipole-dipole scattering configuration in 
Euclidean space. The dipoles have size $a$ and are $b$ apart.
The dipoles are tilted by $\pm\theta/2$ (Euclidean rapidity) in the longitudinal $x_0x_L$ plane. }
\label{FIG_DIPOLES}
\end{figure}

Using the eikonal approximation,
LSZ reduction and the analytic continuation discussed above, 
the dipole-dipole scattering amplitude ${\cal T}$ in Euclidean
space takes the following form~\cite{NACHTMANN}
\be
&&\frac 1{-2is} {\cal T} (\theta , q ) \approx \,\int d^2b\,\, e^{iq_{\perp}\cdot b}\nonumber\\
&&\times
\la \left({\bf W} (-\theta/2, -b/2) -{\bf 1}\right) 
\left( {\bf W} (\theta/2, b/2,) -{\bf 1}\right)\ra\,\nonumber\\
&&=\int d^2b\,\, e^{iq_{\perp}\cdot b}\,\la {\bf W} (-\theta/2, -b/2){\bf W} (\theta/2, b/2) -{\bf 1}\ra \nonumber\\
\label{4X}
\ee
where
\be
{\bf W} (\theta, b)= \frac 1{N_c} {\rm Tr} \left( {\bf P}_c {\rm exp}\left(ig\int_{{\cal C}_{\theta}}\,d\tau\,
A(x)\cdot v\right)\right)\,
\label{5X}
\ee
is the normalized Wilson loop for a dipole, $\langle{\bf W}\rangle\equiv1$. In Euclidean geometry
${\cal C}_{\theta}$  is a closed rectangular loop of width $a$ that is
slopped at an angle $\theta$ with respect to the vertical imaginary time direction (see FIG. \ref{FIG_DIPOLES}).
The two dimensional integral in (\ref{4X}) is over the impact parameter $b$
with $t=-q_{\perp}^2$, and  the averaging is over the gauge configurations
using the QCD action.

In (\ref{4X}-\ref{5X}), the dipole sizes are \textit{fixed}; as such $\cal T$ is their scattering amplitude. In \cite{NACHTMANN}, this amplitude is folded with the target/projectile dipole distributions to generate the pertinent hadron-hadron scattering amplitude.  We note their size $a$ is generic for either longitudinal ($a_L$) or transverse ($a_T$) dipole size. In general, the dipole-dipole scattering amplitude depends on the orientation of the dipoles.  We expect the amplitude to be of the form:
\be
a^2\rightarrow a_T^2+a_L^2/{{\rm sin}^2(\theta/2)}
\label{LT}
\ee
After analytic continuation to Minkowski space, 
the longitudinal orientation is suppressed by a power of $1/s$ which is just the Lorentz contraction
factor. Throughout, $a^2$ will refer to $a_T^2$ as the longitudinal dipole orientation is suppressed
at large $s$.

We will assume that the impact parameter $b$ is large in comparison
to the typical time characteristic of the Coulomb interaction inside
the dipole, i.e. $b\gg \tau_0\approx a/g^2$. As a result the dipoles are
color neutral, and the amplitude in perturbation theory is dominated by
2 gluon exchange. Thus~\cite{SHURYAK}
\be
{\cal T} (\theta , b) \approx \frac {N_c^2-1}{N_c^2}\,\frac {(ga)^4}{32\pi^2}\,
\frac{{\rm cotan}^2\,\theta}{b^4}\,,
\label{6X}
\ee
for two identical dipoles of size $a$ with polarizations along 
the impact parameter $b$. The analytic continuation shows
that ${\rm cotan}\,\theta\rightarrow 1$, leading to a finite total
cross section. We note that ${\cal T}\sim (a/b)^4\lambda/N_c^2$, and thus
subleading at large $N_c$.

\section{Holographic Computation and the Schwinger Mechanism}\label{QQPT}

In this section, diffractive dipole-dipole scattering in holographic QCD will be pursued through
closed string exchanges between the two dipole Wilson loops. 
Instead of working in the semi-classical approximation as originally proposed in~\cite{Rho:1999jm,Janik:2000pp,Janik:2000aj,Janik:2001sc}  and dictated by the tenets of holography, in the present approach we will attempt to compute a full string partition function with reasonable approximations. As a consequence some of our results include subleading $\alpha'$-corrections such as the intercept, although the main focus of our discussion is on the leading large $\lambda$ contributions dominated by semi-classical worldsheets.
Our motivation is to identify these contributions via a more rigorous computation compared to the variational approaches
taken in~\cite{Janik:2000pp,Janik:2000aj,Janik:2001sc}, resolving some of the issues related to the multibranch structures
in them. Also, our computation will give us more physicsal insight on the nature of these semi-classical worldsheets in terms of
a stringy version of the Schwinger mechanism with an electric field induced by the probes relative rapidity.

For small dipoles and large impact parameter $b$, we assume that most of the string worldsheet stays at the IR end point, so that we have effectively a flat-space with an effective string tension neglecting fluctuations along the holographic direction. This approximation is based on the generic form of the confining metric 
\be
ds^2=\frac{dz^2}{z^2 f(z)}+\frac{dx\cdot dx}{z^2}+\cdots\,,
\label{conf_metric}
\ee
where $dx\cdot dx $ is the 4 dimensional flat metric and $\cdots$ stands for an extra compact space depending on a particular string theory compactification which is not important for our argument. For confinement, the function $f(z)$ has a zero at some finite $z=z_0$ in the holographic direction. In order to minimize its area, the string worldsheet connecting the dipoles that are placed on the boundary $z=0$ and separated by a large impact parameter b, rapidly falls down to the IR end-point $z=z_0$. At the horizon where the string lives, the string area is measured in units set by the effective string tension $\sigma_T\equiv{1\over 2\pi \alpha^\prime}={1\over 2\pi l_s^2}{ 1\over z_0^2}$. 
For example, for Witten's~\cite{WITTEN} confining metric we have $\sigma_T={2\over 27\pi} M_{KK}^2 \lambda$.
In fact, this flat-space approximation is valid only in the regime of the soft Pomeron where $(-t)\le M_{KK}^2$~\cite{TAN},
and this will be assumed throughout our paper. 

Also, we will neglect the fermionic degrees of freedom on the string worldsheet, which is a deviating point from the analysis in\cite{TAN}. This is a question of worldsheet one-loop determinant corrections to the leading semi-classical string partition function. 
It is
motivated by the results in~\cite{KINAR} for the standard Wilson loop, where it was 
shown that for the static Wilson loop ($\theta=0$), the worldsheet one-loop contribution to the quark-antiquark Wilson
loop is dominated by massless bosonic degrees of freedom giving a L\"uscher-type contribution, whereby the bosonic mode along holographic direction and all 
worldsheet the fermionic modes become massive and give only subdominant contributions. 
In section \ref{valid}, a more precise condition for this to be valid in our case will be presented, especially in comparison to
the ``locality'' assumption in \cite{TAN} which breaks down for sufficiently large $\chi={\rm ln}\, s$.
Based on these approximations, our problem effectively reduces to the one in the flat space bosonic string theory.
However, when we discuss dipoles of higher representations at the end of the section, the nature of
gauge/gravity correspondence of holographic QCD will be important.

\begin{figure}
\includegraphics[scale=0.85]{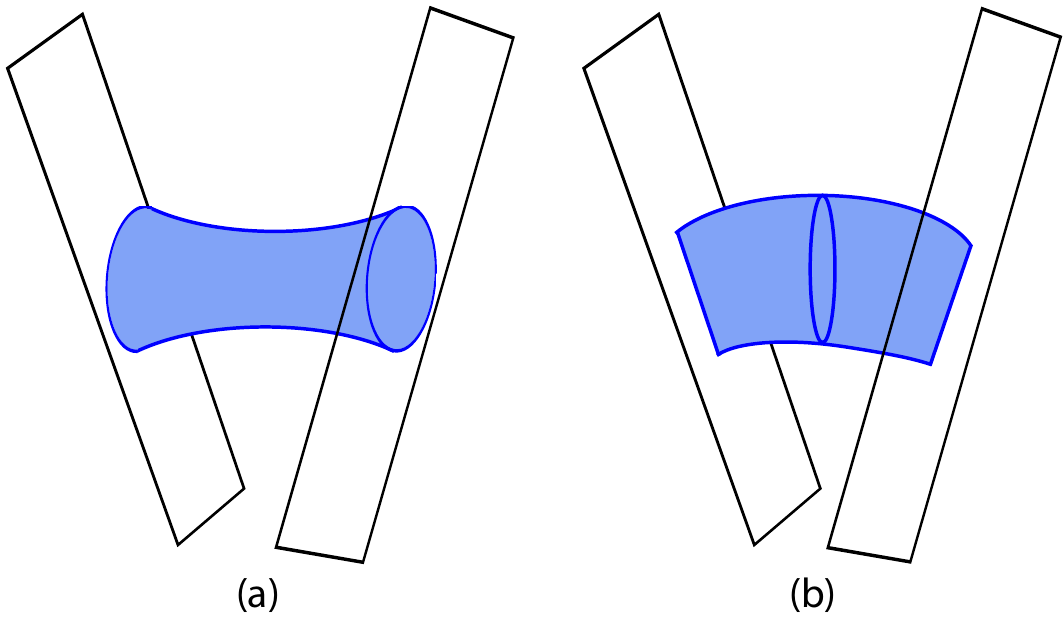}
\caption{(a) Closed string exchange as a funnel contribution (b) Approximation similar to D0 brane scattering with subtle differences explained in the text. The coordinates are the same as in FIG.~\ref{FIG_DIPOLES}.}
\label{2ab}
\end{figure}
The Euclidean connected dipole Wilson loops correlator
\be
{\bf WW}=\la {\bf W} (-\theta/2, -b/2){\bf W} (\theta/2, b/2) -{\bf 1}\ra
\label{CORR}
\ee
appearing in (\ref{4X}) gets the leading large $N_c$ contribution from the exchange of one closed string as in FIG.~\ref{2ab}(a):
the closed string makes a funnel connecting the two dipole Wilson loops.
Note that the funnel has been proposed long time ago as the geometry underlying the Pomeron exchange within the framework of the ``topological expansion" \cite{Veneziano:1976wm}. 
We would like to compute the string partition function summing over all possible fluctuations within the same topology.
This problem is different from the closed string exchange between D-branes in a number of ways:
\begin{enumerate}
\item In our case of funnels, the area inside
the funnel is empty so that the string action is reduced by that amount, whereas for the D-brane case, there is no such effect. 
\item In the D-brane case, multi-winding
of the cylinder topology is allowed without further large $N_c$ suppression, while it is not in the case of emission from a string worldsheet. To have multi-winding, the string genus has to increase leading to further $1\over N_c^2$ suppression. This point will be relevant later when we discuss the truncation of the multi-winding contributions and the dipoles of higher representations. 
\end{enumerate}

As this is a difficult problem in string theory due to a finite dipole size $a$, we necessarily have to make reasonable approximations that would allow us to proceed while still giving us all essential features of the expected result.
For a small dipole size $a$, the two boundaries of the funnel will be highly pinched along the dipole direction, so that
they effectively lie on two straight lines aligned along the direction of the Wilson loop trajectories
as depicted in FIG.~\ref{2ab}(b).
This leads to a reasonable approximation of treating these boundaries strictly sitting on two straight lines inside the two dipole Wilson loops, and the string partition function over these restricted configurations can be computed in a similar
manner as in the case of D0 brane scattering. After that, the locations of the two boundary lines inside each dipole
will be integrated over with a measure naturally obtained from the Polyakov string action, which gives us the final
amplitude with dependency on the dipole size $a$.
As our final result contains all the expected behaviors of Regge trajectory and the intercept, the subset of full configurations
that we have chosen seems to be large enough to contain all essential configurations relevant in the Regge regime.    

As discussed before, there are differences between the real D0 brane scattering amplitude and the amplitude we would like to
compute. The first point  in regard to the area reduction inside the funnels is fine in our approximation because the
D0 brane cannot have a non-zero area anyway. However, the second 
point is relevant and we have to discard all higher winding contributions 
in our final result as they are artifacts of D0 brane and are not the allowed configurations in our original problem. They will
be relevant for scattering of dipoles in higher color representations.

With these in mind, the Euclidean correlator as a string partition function of one closed string exchange is given by
\be
{\bf WW}=g_s^2\int_0^\infty\frac {dT}{2T}\,{\bf K}(T)\,,
\label{SCHWINGER}
\ee
where
\be
{\bf K}(T)=\int_{\rm T} \,d[x]\,e^{-{\bf S}[x]+{\rm ghosts}}\,,
\label{PROP}
\ee
is the string partition function on the cylinder topology with modulus $T$ ($T$ is the circumference of cylinder when its length
is normalized to 1) with suitable boundary conditions that we just discussed above, 
and the Polyakov string action is
\be
{\bf S}={{\sigma}_T\over 2}\int_0^T\,d\tau\int_0^1d\sigma
\left(\dot{x}^\mu\dot{x}_\mu+{x'}^{\mu}{x'}_\mu\right)\,,
\label{POLYAKOV}
\ee
in a conformal gauge $h^{ab}=\delta^{ab}$ for the worldsheet metric.
The ghosts contributions follow from the diagonal gauge-fixing of the metric, and 
for the bosonic string it amounts to two longitudinal ghosts. The dot refers to $\partial_\tau$ and the prime 
refers to $\partial_\sigma$.  
The measure ${dT\over 2T}$ is the well-known measure of conformal classes of worldsheet metrics on the cylinder, and
the factor $g_s^2$ is due to the relative genus in comparison  to the unconnected Wilson loops.
For Witten's geometry, $g_s$ at the IR end point is
\be
g_s={\lambda^{3\over 2}\over \pi 3^{3\over 2} N_c}\quad,
\ee
whose precise form is model-dependent, but the $1/N_c$ suppression is universal.

The integration in (\ref{PROP}) is over periodic configurations
\be
x^\mu(T,\sigma)=x^\mu(0,\sigma)\,,
\label{PERIODIC}
\nonumber
\ee
that stretch between the twisted dipole surfaces in Euclidean space as shown in FIG.~\ref{2ab}(b), with
\be
&&{\rm cos}(\theta/2)\,x^1(\tau, 0)+{\rm sin}(\theta/2)\,x^0(\tau, 0)=0\,,\nonumber\\
&&{\rm cos}(\theta/2)\,x^1(\tau, 1)-{\rm sin}(\theta/2)\,x^0(\tau, 1)=0\,.
\label{TWIST}
\ee
Here, we take the origin in the logitudinal space as the intersection point of the two trajectories at $\sigma=0,1$
projected to the longitudinal space.
As we shift the locations of these two trajectories along the dipole separation width $a$, the intersection point
will move, but the relative geometry of the two trajectories is the same, and it is easily seen that the amplitude
does not depend on these shifts. Therefore, integrating over the dipole size $a$ will simply give us $a^2$
from the two dipoles, up to some unknown constant measure factor that we will discuss later. 
Note that this $a^2$ dependence is a consequence of our approximation of the pinched boundaries for the closed string funnels,
but it will be shown to be more general by the fact that the relevant worldsheet instantons have a small width of order $b/\chi\sim\sqrt{\alpha'/\chi}$ so that it is justified as long as $\sqrt{\alpha'/\chi}\ll a$ for a large $\chi$.
Our result is reminiscent of the dipole parametrization~\cite{GBW}  of the deep-inelastic  
cross section.

The twisted boundary conditions (\ref{TWIST}) are readily  implemented through
\begin{eqnarray}
\begin{pmatrix}
x^0\\ x^1
\end{pmatrix}
=
\begin{pmatrix}
{\rm cos}(\theta_\sigma/2)  &-{\rm sin}(\theta_\sigma/2)
\\ {\rm sin}(\theta_\sigma/2)&{\rm cos}(\theta_\sigma/2)\\
\end{pmatrix}
\begin{pmatrix}
\tilde{x}^0\\ \tilde{x}^1
\end{pmatrix}\,,
\label{XBOOST}
\end{eqnarray}
with $\theta_\sigma=\theta(2\sigma-1)$, and the ordinary Neumann (Dirichlet) boundary condition for $\tilde x^0$ ($\tilde x^1$). The longitudinal $x$ coordinates follow from the $\tilde{x}$ coordinates
by a local rotation on the world-sheet, which implements successive boost transformations on the world-sheet.
(\ref{XBOOST}) is a ruled transformation at the origin of the helicoidal geometry. 
The above transformation is useful because the string fluctuation modes become
purely quadratic in terms of the $\tilde x$ coordinates. The Jacobian of this transformation is 1.

\subsection{Mode Decomposition}

The untwisted coordinates $\tilde{x}^{0,1}$ satisfy both the periodic and usual Neumann/Dirichlet boundary conditions
on the dipole surfaces. The quadratic action in (\ref{PROP}) is easily
diagonalized using
\be
&&\tilde{x}^0(\tau,\sigma)=\sum_{m=-\infty}^{+\infty}\sum_{n=0}^{+\infty}x_{mn}^0\,e^{i2\pi m\tau/T}\,{\rm cos}(\pi n\sigma)\,,\nonumber\\
&&\tilde{x}^1(\tau,\sigma)=
\sum_{m=-\infty}^{+\infty}\sum_{n=1}^{+\infty}x_{mn}^1\,e^{i2\pi m\tau/T}\,{\rm sin}(\pi n\sigma)\,,
\label{LONG}
\ee
which shows how two parallel dipoles with $\theta=0$ (potential problem) get to the twisted dipoles with
$\theta\neq 0$ (scattering problem).  A similar mode decomposition for the potential problem
using the Nambu-Goto string was originally discussed in~\cite{ARVIS}.
The transverse coordinates are untwisted with $x^\perp={\tilde x}^\perp$. They obey both the periodic and Dirichlet boundary conditions. Their mode decomposition is
\be
&&\tilde{x}^\perp (\tau,\sigma)=-b^\perp(1-2\sigma)/2\nonumber\\&&+
\sum_{m=-\infty}^{+\infty}\sum_{n=1}^{+\infty}x_{mn}^\perp \,e^{i2\pi m\tau/T}\,{\rm sin}(\pi n\sigma)\,,
\label{TRANS}
\ee
with the impact parameter $b^\perp$ being two dimensional. Note that the total $x^\perp$ is eight dimensional.

We note that $\tilde{x}^{0}$ includes zero modes which are constant over the $\sigma$ coordinate,
\be
&&\tilde{x}_{ZM}^0(\tau,\sigma)\equiv\sum_{m=-\infty}^{+\infty}e^{i2\pi m\tau/T}
x_{m0}^0\,.
\label{ZMLONG}
\ee
They induce the zero modes in the original coordinates as
\begin{eqnarray}
{\begin{pmatrix}
x^0\\ x^1
\end{pmatrix}}_{ZM}
=
\begin{pmatrix}
{\rm cos}(\theta_\sigma/2)  &-{\rm sin}(\theta_\sigma/2)
\\ {\rm sin}(\theta_\sigma/2)&{\rm cos}(\theta_\sigma/2)\\
\end{pmatrix}
{\begin{pmatrix}
\tilde{x}^0\\ 0
\end{pmatrix}}_{ZM}\,,
\end{eqnarray}
which is a periodic form of the helicoidal surface.
We now use the mode decompositions (\ref{LONG}) and (\ref{TRANS}) to compute the string propagator (\ref{PROP}).

\subsection{${\bf K}(T)$}

Since the Polyakov action is quadratic with the above mode expansions on the world-sheet, it can be factored out 
into its basic contributors,
\be
{\bf K}={\bf K}_{OL}\times{\bf K}_{\O L}\times{\bf K}_{\rm ghost}\times{\bf K}_\perp\,,
\label{ALL}
\ee
where ${\bf K}_{OL}$, ${\bf K}_{\O L}$  represent the longitudinal zero and non-zero mode contributions respectively. ${\bf K}_\perp$ is the transverse mode contribution, and ${\bf K}_{\rm ghost}$ is the extra ghost contribution required by the covariant gauge fixing.
Below we will provide a detailed description of the calculation of ${\bf K}_\perp$ using standard zeta function regularization. The other contributions 
follow similarly, and will only be quoted as final results.

The transverse mode decomposition (\ref{TRANS}) once inserted in (\ref{POLYAKOV}) yields 
products of Gaussian integrals for the transverse modes
\be
{\bf K}_\perp(T)&=&e^{-\sigma_T b^2T/2}
\prod_{n=1}^{+\infty}\prod_{m=-\infty}^{+\infty}\int_{-\infty}^{\infty}dx^\perp_{mn}\nonumber\\
&&\times \exp\left(-\frac{\sigma_T \,\pi^2}{4}\left(\frac{4 m^2}{T}+n^2 T\right)x^{\perp^2}_{mn} \right)\nonumber\\
&=&e^{-\sigma_T b^2T/2}
\prod_{n=1}^{+\infty}\prod_{m=-\infty}^{+\infty}\nonumber\\
&&\times\left(\frac{\sigma_T \, \pi}{4T}(4m^2+n^2 T^2)\right)^{-{D_\perp}/{2}}\,.
\label{KPERP}
\ee
The infinite products in (\ref{KPERP}) can be evaluated by using zeta function regularization technique. Indeed, the
infinite product of a constant can be written as
\be
\prod_{n=1}^{+\infty}c=e^{\ln c \sum_{n=1}^{\infty}\,1}=e^{\zeta(0)\, \ln c} =c^{-1/2}\,,
\ee
where $\zeta(s)=\sum_{n=1}^{\infty}n^{-s}$ is the Riemann zeta function. In particular this leads to
\be
\prod_{n=-\infty}^{+\infty}c=1\,,
\ee
since the zero mode and non-zero mode contributions cancel. Similarly from analytic continuations $\zeta^\prime(0)=-{1/ 2}\ln(2 \pi)$ and $\zeta(-1)=\-1/12$ we get
\be
\prod_{n=1}^{+\infty}n=\sqrt{2\pi}\quad,\quad\prod_{n=1}^{+\infty}e^{-a\,n}&=&e^{a/12}\,.
\label{ZETA}
\ee
Finally by using the product formula for $\sinh$
\be
\sinh(\pi\,x)=\pi\,x \prod_{n=1}^{+\infty}\left(1+\frac{x^2}{n^2}\right)\,,
\ee
the transverse contribution $\bf K_\perp$ can be put into the form,
\be
{\bf K}_\perp(T)=e^{-\sigma_T b^2T/2}
\prod_{n=1}^{+\infty}\left(2 \sinh ({n \pi}T/{2})\right)^{-D_\perp}\,,
\ee
and the second identity in (\ref{ZETA}) can be used to express ${\bf K}_\perp(T)$ in the standard form
\be
{\bf K}_\perp(T)&=&e^{-\sigma_T b^2T/2}
\eta^{-D_\perp}(iT/2)\,,
\label{KT}
\ee
where $\eta$ is the Dedekind eta function,
\be
\eta(\tau)\equiv q^{1/24}\prod_{n=1}^{+\infty}\left(1-q^n\right)\quad,\quad q=e^{2i \pi \tau}\,.
\ee

The longitudinal mode contribution to the string propagator follows similarly by inserting (\ref{LONG}) in
the Polyakov action (\ref{POLYAKOV}) and carrying out the Gaussian integration. This contribution can be
separated into the zero mode contribution as given by (\ref{ZMLONG}) and the non-zero mode longitudinal
contribution. Specifically, the longitudinal zero mode part contributes
\be
&&{\bf K}_{OL}(T)=\prod_{m=-\infty}^{+\infty}\left(\frac{4 \pi^2 m^2}{T}+\theta^2 T \right)^{-1/2} \nonumber\\
&&=\frac{1}{2\sinh (\theta T/2)}\,,
\label{KZML}
\ee
while the longitudinal non-zero mode part contributes
\be
{\bf K}_{\O L}(T)=&&\prod_{m=-\infty}^{+\infty}\prod_{n=1}^{+\infty}\prod_{s=\pm}\nonumber\\
&&\times \left(\frac{4 m^2}{T}+\left(n+\frac{s\theta}\pi\right)^2T\right)^{-1/2}\nonumber\\
=&&\prod_{n=1}^{+\infty}\prod_{s=\pm}\frac{1}{2\sinh\left((n+s\frac{\theta}{\pi})\frac{\pi T}{2}\right)}\,.
\label{KNZML}
\ee
Notice that for the longitudinal modes, $\theta/\pi$ plays the role of a Bohm-Aharonov phase
that modifies the azimuthal quantum number $n$. This observation shows
that the sloping or twisting of the Wilson lines in Euclidean space is dual to an "electric/magnetic"
field in the longitudinal directions (electric and magnetic fields are indistinguishable in Euclidean space). 
It is this longitudinal electric field that is at the origin of the Schwinger mechanism.
A more direct way of seeing the Schwinger mechanism via worldsheet T-duality and its worldsheet instantons will be explained shortly.

The ghost contribution tags to the two longitudinal non-zero mode contributions and is unaffected by
the twist. Its contribution to (\ref{ALL}) is
 \be
{\bf K}_{\rm ghost}(T)&=&\prod_{m=-\infty}^{+\infty}\prod_{n=1}^{+\infty}
\left(\frac{4m^2}{T}+n^2T\right)^{+1}\nonumber\\
&=&\prod_{n=1}^{+\infty}4\sinh^2(n \pi T/2)\,.
\label{KGHOST}
\ee
Combining all the terms (\ref{KT}-\ref{KGHOST}) in (\ref{ALL}) leads to the full periodic propagator
\be
&&{\bf K}(T)=\frac{a^2/\alpha'}{2\,{\rm sinh}(\theta T/2)}\nonumber\\ &&\times \prod_{n=1}^{+\infty}\prod_{s=\pm}
\frac{{\rm sinh}\,(\pi nT/2)}{{\rm sinh}(\pi (n+s\theta/\pi)T/2)}\nonumber\\
&&\times e^{-\sigma_T b^2 T/2}\eta^{-D_\perp}(iT/2)\,,
\label{FINAL}
\ee
where we include the factor $a^2$ from integrating over the dipole width $a$.
The first contribution in (\ref{FINAL}) stems from the longitudinal zero modes, 
the second contribution is from the longitudinal non-zero modes including
the ghost fields, and the final contribution arises from the $D_\perp$ transverse
modes. 

By dimensional reasoning, there must be a $1/\alpha'$ multiplying $a^2$ which should come from the integration measure.
The overall unknown numeric constant of this measure can be reabsorbed into our definition of the dipole parameter $a$.

\subsection{$\bf WW$}

The contribution of (\ref{FINAL}) to the elastic dipole-dipole amplitude at fixed impact parameter
follows from inserting it into (\ref{CORR}). For small dipoles for which our approximations are justified, 
${\bf WW}$ takes the  form
\be
{\bf WW}= g_s^2 \int_0^\infty\,\frac{dT}{2T}\,{\bf K}(T)\,,
\label{NOUNI}
\ee
which shows that the elastic amplitude vanishes as the dipole size $a\rightarrow 0$. The 
phenomenological relevance of (\ref{NOUNI}) to deep-inelastic scattering including the possible connection to the saturation phenomena will be discussed elsewhere.

Using (\ref{FINAL})
after the analytic continuation to Minkowski space $\theta\rightarrow -i\chi$ gives
\be
&&{\bf WW}= \frac {ig_s^2a^2}{4\alpha'}\int_0^\infty\frac{dT}{T}
\frac{1}{\,{\rm sin}(\chi T/2)}\nonumber\\ &&\times \prod_{n=1}^{+\infty}\prod_{s=\pm}
\frac{{\rm sinh}(\pi nT/2)}{{\rm sinh}(\pi (n+is\chi/\pi)T/2)}\nonumber\\
&&\times e^{-b^2 T/4\pi\alpha'}\eta^{-D_\perp}(iT/2)\,.
\label{MINFINAL}
\ee
The zero mode contribution in (\ref{MINFINAL}) developes poles along the real T-axis
for ${\rm sin}(\chi T/2)=0$ or $T=2k\pi/\chi$.  Feynman prescriptions for the elastic scattering 
amplitude requires deforming the contour above the negative poles and below the positive
poles. For $\chi\rightarrow \infty$, the contribution at the poles is purely real
\be
{\bf WW}_{poles}=&&\frac {g_s^2a^2}{4\alpha'}
\sum_{k=1}^\infty\frac{(-1)^{k}}k\,\eta^{-D_\perp}(ik\pi/\chi)\nonumber\\
&&\times e^{-k b^2/2\chi\alpha'}
\label{POLES}\,,
\ee
which gives the inelasticity that we are interested in.
We note that (\ref{POLES}) displays an essential singularity 
as $\chi\rightarrow 0$, which is a hallmark of tunneling. This is related to the fact that they are generated through
worldsheet instantons via the  Schwinger mechanism as we detail below.

Since we are interested in the $\chi\to\infty$ limit, the above expression, which is written in the open string viewpoint, is not
suitable to correctly identify the limit, and one needs to transform it to a closed string viewpoint 
by using the modular relation of the Dedekind eta function, $\eta(i\,x)=\eta(i/x)/\sqrt{x}$ \cite{APOSTOL}. We have
\be
{\eta^{-D_\perp}(ik\pi/\chi)}=&&\left(\frac{k\pi}{\chi}\right)^{D_\perp/2}e^{D_\perp\chi/12k}\nonumber\\
&&\times\prod_{n=1}^{+\infty}\left(1-e^{-2\chi n/k}\right)^{-D_\perp}\,.
\label{DEDEX}
\ee
Also
\be
\prod_{n=1}^{+\infty}\left(1-e^{-2\chi n/k}\right)^{-D_\perp}=\sum_{n=0}^{\infty}d(n)\,e^{-2\chi\,n/k}\,,
\label{DENSITY}
\ee
exhibits a harmonic spectrum. It is the generating function of the bosonic string level density. Asymptotically~\cite{FUBINI}
\be
d(n)\approx \frac{e^{2\pi\,\sqrt{D_\perp\,n/6}}}{n^{D_\perp/4}}\,.
\label{DENSITY1}
\ee
The exponentially rising density (\ref{DENSITY1}) is a hallmark of string excitations. We note that $d(0)=1$.

\subsection{Schwinger mechanism}

In this section, we will provide a physical understanding of the non-perturbative contributions in the exponent of (\ref{POLES}), 
that is the terms ${\rm exp}(-k b^2/2\chi\alpha')$ which drive the Regge behavior in momentum space for $(-t)>0$ as we will see in section \ref{ELAS}. Recall that the $k$'th contribution comes from the pole at $T=T_k=2\pi k/\chi$, which will be important later.
The nature of these contributions indicates that they should arise from semi-classical worldsheet instantons. These
world-sheet instantons bear some similarities to the instantons/sphalerons advocated in 
\cite{SHURYAK,KHARZEEV} to play an important role in diffractive scattering.

We will shed more light on this by showing that upon worldsheet T-duality these worldsheet instantons map to a
stringy version of well-known instantons in the Schwinger mechanism
of pair creation under external electric field, where in our case the electric fields acting on the end points of the open string
is triggered by  the relative rapidity of the probes via T-duality. 
We will find the analytic solutions for these worldsheet instantons in the stringy Schwinger mechanism, and show that the $k$'th contribution arises from a $k$-wrapping worldsheet instanton solution, much like point particles.  Moreover, we
will show that these tunneling configurations last a time $T=T_k=2\pi k/\chi$ and carry an on-shell action
$S_k=k b^2/2\chi\alpha'$.

We start from our assumption that the two boundaries of the cylinder worldsheet sit on the straight lines with rapidity angles 
$\chi/2$ and $-\chi/2$ for $\sigma=0,1$ respectively. These are effectively the same boundary conditions as in the D0 brane 
scattering set up. At $\sigma=0$ (the analysis for $\sigma=1$ will be similar and we will simply present the final result later)
the boundary condition can be written explicitly as
\be
&&\cosh(\chi/2)\partial_\sigma x^0 +\sinh(\chi/2)\partial_\sigma x^1=0\,,\nonumber\\
&&\sinh(\chi/2)\partial_\tau x^0+\cosh(\chi/2)\partial_\tau x^1=0\,.\label{bdr1}
\ee
We then invoke a worldsheet T-duality along the direction $x^1$,
\be
\partial_\tau x^1=\partial_\sigma y^1\quad,\quad\partial_\sigma x^1=\partial_\tau y^1\,,
\ee
to have a dual description in terms of $y^1$.
Note that the worldsheet instantons we will present shortly are in the
zero winding/momentum sector, so that the compactification of  the $x^1$ direction and its radius transformation in T-duality 
is not relevant for our purposes. This is a technical tool to find the worldsheet instantons in the original problem.
The boundary condition (\ref{bdr1}) then becomes
\be
&&\cosh(\chi/2)\partial_\sigma x^0 +\sinh(\chi/2)\partial_\tau y^1=0\,,\nonumber\\
&&\sinh(\chi/2)\partial_\tau x^0+\cosh(\chi/2)\partial_\sigma y^1=0\,,\label{bdr2}
\ee
which is easily shown to be equivalent to putting a boundary term to the Polyakov action,
\be
S&=&{\sigma_T\over 2}\int d\tau\int d\sigma \left(-(\partial x^0)^2+(\partial y^1)^2+(\partial x^\perp)^2\right) \nonumber\\
&+&{E\over 2}\int d\tau \left(y^1\partial_\tau x^0-x^0\partial_\tau y^1\right)\bigg|_{\sigma=0,1}\,,
\ee
with 
\be
E=\sigma_T \tanh(\chi/2)\,,\label{echi}
\ee
being an electric field along the $y^1$ direction, $F_{10}=E$. This aspect is a well-known feature of T-duality in D-brane physics. Note that the electric field acts on the two end points of the open strings stretching between the two dipoles. The signs of the electric fields on both ends are opposite due to the opposite direction of motions of the two dipoles, but the two end points of a string carry opposite charges, so that there is a net acceleration. This explains the existence of a Schwinger mechanism of
pair creation of strings in high energy collisions.
 
To find the worldsheet instantons of this stringy version of the Schwinger mechanism, we 
proceed  to the Euclidean description with an action
\be
S_E= 
&=&{\sigma_T\over 2}\int_0^T d\tau\int_0^1 d\sigma \left((\partial x^0)^2+(\partial y^1)^2+(\partial x^\perp)^2\right) \nonumber\\
&+&{E\over 2}\int_0^T d\tau \left(y^1\partial_\tau x^0-x^0\partial_\tau y^1\right)\bigg|_{\sigma=0,1}\nonumber\\
&=&{\sigma_T\over 2}\left(I_\sigma T +{I_u\over T}\right)\nonumber\\
&+&{E\over 2}\int_0^1 du \left(y^1\partial_u x^0-x^0\partial_u y^1\right)\bigg|_{\sigma=0,1}\,,\label{se}
\ee
where we have changed the variable $\tau\equiv T u$, and $I_{\sigma,u}$ are defined by
\be
I_\sigma&\equiv&\int_0^1 d\sigma\int_0^1 du\,\left((\partial_\sigma x^0)^2+(\partial_\sigma y^1)^2+(\partial_\sigma x^\perp)^2\right)\,,\nonumber\\
I_u&\equiv&\int_0^1 d\sigma\int_0^1 du\,\left((\partial_\tau x^0)^2+(\partial_\tau y^1)^2+(\partial_\tau x^\perp)^2\right)\,.\nonumber\\\label{Iu}
\ee
We have to find saddle the points of the action (\ref{se}) with respect to both the $T$-integral and the worldsheet fields $(x^0,y^1,x^\perp)$, based on the largeness of $\sigma_T\sim E\sim\lambda$.

A similar problem was solved in~\cite{SCHUBERT}, and we follow the same steps to find the explicit solutions.
The $T$-dependence is algebraic, and it is easy to find its saddle point as
\be
T=\sqrt{I_u/I_\sigma}\,.\label{Tsaddle}
\ee
Inserting this back into (\ref{se}) gives 
\be
S_E=\sigma_T\sqrt{I_u I_\sigma}+{E\over 2}\int_0^1 du \left(y^1\partial_u x^0-x^0\partial_u y^1\right)\bigg|_{\sigma=0,1},\nonumber\\\label{se2}
\ee
whose equations of motion are
\be
\partial_\sigma^2(x^0,y^1,x^\perp)+{I_\sigma\over I_u}\partial_u^2(x^0,y^1,x^\perp)=0\,,
\ee
with the boundary conditions
\be
\sqrt{I_u\over I_\sigma}\partial_\sigma(x^0,y^1)\pm(-1)^\sigma{E\over\sigma_T}\partial_u (y^1,x^0)\bigg|_{\sigma=0,1}=0\,.
\ee
The Dirichlet boundary condition for $x^\perp$ fixes its solution as $x^\perp=b \sigma$, and
for $(x^0,y^1)$ we write a $k$-wrapping Ansatz as
\be
x^0=R(\sigma)\cos(2\pi k u)\quad,\quad y^1=R(\sigma)\sin(2\pi k u)\,,\label{ansatz}
\ee
with a $\sigma$-dependent radius function $R(\sigma)$ to be determined. 
With (\ref{ansatz}), we have
\be
I_u=(2\pi k)^2\int_0^1 d\sigma\, (R(\sigma))^2\,,\, I_\sigma=b^2+\int_0^1 d\sigma (R'(\sigma))^2\,,\nonumber\\\label{Iu2}
\ee
and the equation of motion for $R(\sigma)$ is
\be
R''(\sigma)-(2\pi k)^2{I_\sigma\over I_u} R(\sigma)=0\,,\label{eq}
\ee
with the boundary condition
\be
\sqrt{I_u\over I_\sigma}R'(\sigma)+(-1)^\sigma {2\pi k E\over\sigma_T} R(\sigma)\bigg|_{\sigma=0,1}=0\,.\label{bd2}
\ee
The unique consistent solution of (\ref{Iu2}), (\ref{eq}), and (\ref{bd2}) is possible only for $k>0$, and is given by
\be
R(\sigma)&=&{b\over 2{\rm arctanh}(E/\sigma_T)}\cosh\left({\rm arctanh}(E/\sigma_T)\left(2\sigma-1\right)\right),\nonumber\\
&=&{b\over\chi}\cosh\left(\chi\left(\sigma-1/2\right)\right)\,,
\label{Rsol}
\ee
with
\be
\sqrt{I_u\over I_\sigma}={2\pi k\over 2{\rm arctanh}(E/\sigma_T)}={2\pi k/\chi}\,,
\ee
where we have used (\ref{echi}), $E=\sigma_T \tanh(\chi/2)$.
From (\ref{Tsaddle}) we see that this corresponds to $T=2\pi k/\chi$ confirming our expectation. 
The value of the on-shell action $S_E$ in (\ref{se}) is also easily computed as 
\be
S_E=\sigma_T (2\pi k){b^2\over 2\chi}={k b^2\over 2 \chi\alpha'}=S_k\,,
\ee
which precisely agrees with the negative of the exponent in ${\rm exp}(-k b^2/2\chi\alpha')$.
These are convincing evidences that the Regge behavior of soft Pomeron exchange is indeed driven by a
Schwinger mechanism
of pair creating strings, where the effective electric fields are induced by the rapidity of the projectiles.

The circular motion of the string instanton on the Euclidean longitudinal plane with the $\sigma$-dependent radius $R(\sigma)$ becomes 
an accelerating hyperbolic motion in Minkowski spacetime. The resulting $\sigma$-dependent acceleration $a(\sigma)$ is
\be
a(\sigma)=1/R(\sigma)={\chi\over b}{1\over\cosh\left(\chi\left(\sigma-1/2\right)\right)}\,,
\ee
which has a maximum at the center of the string, $\sigma=1/2$. Due to this acceleration, the string feels
a $\sigma$-dependent Unruh temperature,
\be
T(\sigma)={a(\sigma)\over 2\pi}={\chi\over 2\pi b}{1\over\cosh\left(\chi\left(\sigma-1/2\right)\right)}\,,
\ee
with a maximum $T_m={\chi\over 2\pi b}$ at the center.
The temperature quickly drops to a small value around the two end points. 
The existence of this finite temperature may naturally explain the diffusion-like phenomena
noted  in Pomeron physics, the details of which will be discussed elsewhere.

It is very interesting to compare the Unruh temperature $T(\sigma)$ with the Hagedorn temperature $T_H$ and/or deconfinement
transition temperature $T_D$ which characterize the transition temperature to a plasma phase.
The effective Hagedorn temperature is given by
\be
T_H=\sqrt{3\over 2\pi^2 D_\perp \alpha'}=\sqrt{3\sigma_T\over \pi D_\perp}={\sqrt{2}\over 3\pi}\sqrt{\lambda\over D_\perp} M_{KK}\,,
\ee
where the last expression is for the Witten's geometry.
The deconfinement temperature of the same model is $T_D={M_{KK}\over 2\pi}$~\cite{Aharony:2006da}, thus $T_H\ge T_D$
at strong coupling.
In section \ref{ELAS} where we go to the momentum space Regge behavior, we will see that the dominant contribution
in the $b$-integral for a fixed $\sqrt{-t}$ comes from a region where
\be
b\sim {\rm min}\left(\sqrt{2\chi\alpha'}, {1\over\sqrt{-t}}\right)
\,,\label{bchi}
\ee
so that we have two different cases:
\begin{enumerate}
 \item When $\sqrt{-t}\le M_{KK}\sqrt{2\lambda/27\chi}$, we have $b\sim \sqrt{2\chi\alpha'}$, so
$T_m\ge T_D$, i.e. the middle region of the string feels a temperature greater than the deconfinement temperature when 
\be
\chi\ge {27\over 2\lambda}\,.\label{cond1}
\ee

\item In the other case of $\sqrt{-t}\ge M_{KK}\sqrt{2\lambda/27\chi}$, we have $b\sim 1/\sqrt{-t}$, and $T_m\ge T_D$ when
\be
\chi\ge M_{KK}/\sqrt{-t}\,.\label{cond2}
\ee
\end{enumerate}
Note that our soft Pomeron picture is valid when $\sqrt{-t}\le M_{KK}$, so that $\chi \ge {\cal O}(1)$ can easily satisfy both (\ref{cond1}) and (\ref{cond2}).

\begin{figure}
\includegraphics[scale=1]{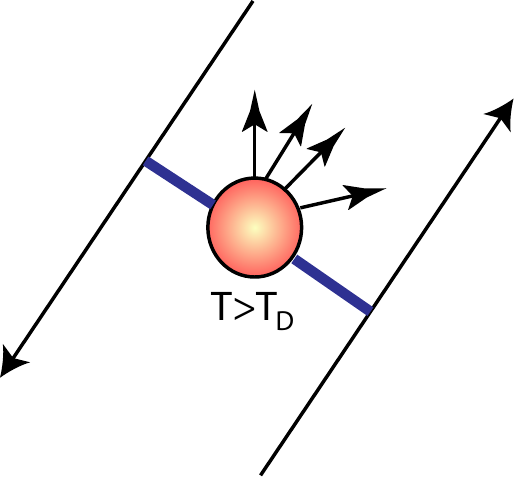}
\caption{In the middle of the transverse string worldsheet, there exists a micro-fireball with an effective Unruh temperature that exceeds the deconfinement temperature. }
\label{Micro}
\end{figure}
To summarize, in the middle of the created string there is a small region where the temperature is higher than the deconfinement temperature and the string description should be replaced by a plasma phase. As the temperature quickly drops away from the center, the plasma size is small: we call it ``micro-fireball''. See FIG. \ref{Micro}. A simple computation gives its transverse size $\Delta x^\perp=b\Delta \sigma$ as
\be
\Delta x^\perp \sim \sqrt{27/2\lambda\chi} M_{KK}^{-1}{\rm ln}\left(8\lambda\chi\over 27\right)\,,
\ee
for the case 1, and
\be
\Delta x^\perp\sim {2\over\chi\sqrt{-t}}{\rm ln}\left(2\chi\sqrt{-t}\right)\,,
\ee
for the case 2. For large $\chi\to\infty$, both become small.

This is an important observation. The existence of a micro-fireball from a single soft Pomeron exchange
can naturally explain the observed apparent thermal nature of multiparticle production in high energy collisions. The Unruh radiation in QCD was previously argued to be responsible for the apparent thermalization in Refs. \cite{Kharzeev:2005iz,Kharzeev:2006zm,Castorina:2007eb,Zhitnitsky:2012im}. This phenomenon may also give a new insight 
on the origin of the diffusion-like behavior in the impact parameter space (``Gribov diffusion") and in the  
transverse momentum space.
The micro-fireball is a consequence of the non-perturbative aspects of
QCD with soft Pomerons, which is not clearly seen in the regime of perturbative QCD. 

This point can be made more transparent after  inserting  (\ref{DEDEX}-\ref{DENSITY1}) into (\ref{POLES}),
leading
\be
{\bf WW}_{poles}&&=\frac {g_s^2 a^2}{4\alpha'}
\sum_{k=1}^\infty\sum_{n=0}^\infty \frac{(-1)^{k}}k \left(\frac{k\pi}{\chi}\right)^{D_\perp/2}\nonumber\\
&&\times d(n) e^{-kb^2/2\chi\alpha'+D_\perp\chi/12k-2\chi n/k}\,.
\label{XPOLES}
\ee
and noting that
\be
{\bf K}(\chi, b) =\left(\frac{k}{2\pi\alpha'\,\chi}\right)^{D_\perp/2}\,e^{-kb^2/2\chi\alpha'}\,,
\label{DIFF}
\ee
is the normalized diffusion propagator in $D_\perp$ dimensions,
\be
\partial_\chi\,{\bf K}(\chi, b) ={\bf D}\,\nabla^2_\perp\,{\bf K}(\chi, b)\,.
\label{DIFF1}
\ee
The diffusion constant in rapidity space is ${\bf D}=\alpha'/2k$.  For long strings, the diffusion propagator (\ref{DIFF}) 
emerges as the natural version of the periodic string propagator in (\ref{SCHWINGER}-\ref{PROP}) in the diffusive regime $b\sim \sqrt{\chi\alpha'}$.

\subsection{Truncation of the $k$-sum and dipoles of higher representations}

Since (\ref{XPOLES}) is ultimatly tied with the total cross section in impact parameter space
as in (\ref{CROSS}) below, it behooves us to interpret the appearance of the $(-1)^k$ in the k-ality sum. 
Schematically, the sum can be written as
\be
\sum_{k=1,3,5,..}e^{-S_k}
-\sum_{k=2,4,6,..}e^{-S_k}\,.
\label{EVENODD}
\ee
So the odd k-ality sum yields instantons, while the even k-ality sum yields antinstantons. Indeed, the
instantons produce pair of close strings by tunneling {\it forward} while the antinstantons annihilate
pair of close strings by tunneling {\it backward}. This back-and-forth process is allowed because there
is no constaint on the bosonic pair creation process in the Schwinger mechanism. This is not true for the
fermionic pair creation process. Incidentally, this back-and-forth process reminiscent of instanton-anti-instanton
dynamics may be the world-sheet analogue of the sphaleron mechanism suggested in~\cite{SHURYAK}.
Indeed, standard instantons contribute $e^{-S_k}$ to the tunneling amplitude and $e^{-2S_k}$ to the probability,
prompting us to rewrite the exponents in (\ref{EVENODD}) as $\sqrt{e^{-2S_k}}$ which is a sphaleron
probability.

It is clear that the $k$'th contribution comes from the semi-classical
worldsheet which wraps the $k=1$ configuration $k$ times. Although these multi-winding contributions are perfectly fine
in the case of a real D0 brane scattering, they are in fact not allowed topologically in our case of closed string emission/absorption from the string worldsheets of two dipole Wilson loops.
To understand this we note that we are originally summing over funnels, and having multiple funnels on top of each other
changes the genus of the total string worldsheet, which entails further $1\over N_c^2$ suppressions.
This means that only the $k=1$ contribution in (\ref{XPOLES}) is physical while higher winding $k>1$ contributions are artifacts of our D0 brane analogue. It is interesting that a similar kind of ambiguity appeared in the variational approach
in~\cite{Janik:2000aj,Janik:2000pp,Janik:2001sc}, where one gets similar $k>1$ contributions from the multibranch structure of the
minimized Nambu-Goto action. Our discussion indicates that these multibranch contributions arise from worldsheet configurations that are prohibited by topology without further $1\over N_c^2$ suppression, and hence should be discarded at large $N_c$.

\begin{figure}
\includegraphics[scale=0.6]{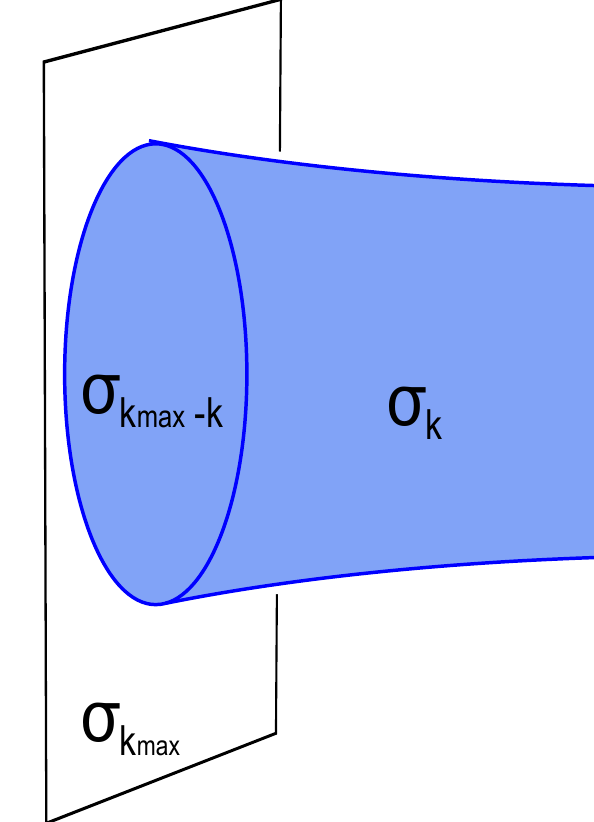}
\caption{If the dipoles belong to a representation with central charge $k_{max}$, then they can exchange strings with k-ality $k\le k_{max}$. Charge conservation dictates that the part of the string worldsheet inside the dipole has k-ality $k_{max}-k$. }
\label{FIG_TENSION}
\end{figure}
Although the above conclusion is true for the Wilson loops in the fundamental representation, 
the situation can change if one considers dipole Wilson loops of higher representations.
Intuitively it is clear that the worldsheet that a Wilson loop of higher representation bounds should be a composite object made of multiple overlapping fundamental string worldsheets.
When the representation is constructed from a product of $k$ fundamental representations,
the corresponding worldsheet that bounds the Wilson loop should be a composite object made of $k$ fundamental strings. 
When $k\ll N_c$, the distinction between this object and the simple non-interacting $k$ fundamental strings is small,
whereas for $k\sim N_c$ the composite object is quite different from the simple sum of fundamental strings, and it is typically described by D-branes wrapping appropriate cycles. For example, in Witten's geometry, the $k$-antisymmetrized representation, corresponding to $k$-string, is described by D4 brane wrapping the internal $S^3\subset S^4$ cycle, whose string tension features Casimir scaling~\cite{Callan:1999zf}
\be
\sigma_k=\sigma_T k(N_c-k)/(N_c-1)\,,\label{kten}
\ee
although the precise form of the string tension $\sigma_k$ is model-dependent~\cite{SINELAW}.

On these composite worldsheets made of $k_{\rm max}$ fundamental strings, it is indeed possible to attach
$k$ multi-winding worldsheets of fundamental strings up to $k\le k_{\rm max}$.
It is easy to understand this as in FIG. \ref{FIG_TENSION}. For example, if dipole the Wilson loops 
in the $k_{\rm max}$-antisymmetrized representation emit/absorb $k$ multi-wound strings, the interior of the funnel should be a
$(k_{\rm max}-k)$-string worldsheet
by string charge conservation. This gives an inequality $k\le k_{\rm max}$. 
Therefore, in the sum (\ref{XPOLES}) one might keep the terms up to $k\le k_{\rm max}$.

However, there are two subtleties regarding this. The first one is the additional large $N_c$ suppression as $k$ becomes close to $k_{\rm max}$. 
The way to count the $N_c$ dependence is the following. One can think of a $k_{\rm max}$-string as a simple sum of a $k_{\rm max}$ number of fundamental strings for the purpose of large $N_c$ counting.
Assume that one fundamental string gets emitted 
from them. The emission from a single string entails $g_s\sim {1\over N_c}$, and there
are $k_{\rm max}$ possible ways to attach the emitted string, so this process has $k_{\rm max}\over N_c$ factor
as a coupling constant.  For the two string emission
(corresponding to $k=2$), one has $k_{\rm max}(k_{\rm max}-1)\over 2N_c^2$ because a single string cannot emit two strings without a large $N_c$ suppression. For a general $k$, it is $_{k_{\rm max}}C_k\cdot N_c^{-k}$. When $k_{\rm max}\sim N_c$, there is indeed no additional large $N_c$ suppression in the summation over $k$ for small $k$, but when $k\sim k_{\rm max}$ it is clear that they are affected by an additional large $N_c$
suppression. 
Another subtlety is that the $k$'th contribution in (\ref{XPOLES}) contains the tension of $k$ number of strings as $k\sigma_T$, which can be seen in the first term in the exponent of the second line.
When $k\sim N_c$ in the case of $k_{\rm max}\sim N_c$, this tension should be replaced by the suitable $k$-string tension, for example (\ref{kten}). As a result, one can really trust the $k$-sum in (\ref{XPOLES}) only for small $k\ll N_c$.

\section{Holography: Elastic Amplitude}\label{ELAS}

The elastic dipole-dipole scattering amplitude follows from (\ref{4X}) after inserting the pole contributions
(\ref{XPOLES}). Performing the integration over transverse $b$ yields  
\be
\frac 1{-2is} {{\cal T}(s,t)} \approx &&\frac {\pi^2 g_s^2 a^2}{2}
\sum_{k=1}^{k_{\rm max}}\sum_{n=0}^\infty\frac{(-1)^{k}}k \left(\frac{k\pi}{{\rm ln}\,s}\right)^{D_\perp/2-1}\nonumber\\
&&\times d(n)\,s^{-2n/k+D_\perp/12k+\alpha't/2k}\,,
\label{DIPOLEAMPLITUDE}
\ee
with $k_{\rm max}$ depending on the representation. 
Although the Gaussian $b$-integral is dominated by the imaginary saddle point
\be
b=i \alpha'\chi\sqrt{-t}/k\,,\label{bsaddle}
\ee
in the real $b$-space it is clear that the dominant region is
\be
b\sim {\rm min}\left(\sqrt{2\chi\alpha'/k}, {1\over\sqrt{-t}}\right)\,.\label{bsaddle}
\ee
All the $n\neq 0$ contributions from string vibrations
are suppressed by $s^{-2n/k}$ relative to $n=0$ contributions at large $s$. Thus
\be
\frac 1{-2is} {{\cal T}(s,t)} \approx \frac {\pi^2 g_s^2 a^2}{2}
\sum_{k=1}^{k_{\rm max}}\frac{(-1)^{k}}k \left(\frac{k\pi}{{\rm ln}\,s}\right)^{D_\perp/2-1}s^{\alpha_{{\bf P}k}(t)-1}\nonumber\\
\label{DIPOLEAMPLITUDE}
\ee
where
\be
{\alpha_{{\bf P}k}(t)}=1+\frac{D_\perp}{12k}+\frac{\alpha'}{2k}t\,.
\label{SLOPE1}
\ee
Therefore we have multiple Pomeron-like trajectories of ${\alpha_{{\bf P}k}(t)}$.
One has ${\alpha_{{\bf P}k}(t)}>{\alpha_{{\bf P}(k+1)}(t)}$ when 
\be
(-t)<{D_\perp\over 6\alpha'}={\pi D_\perp\sigma_T\over 3}={2 D_\perp\over 81}M_{KK}^2\lambda\,,
\ee
which is always satisfied for the soft Pomeron regime, so that the leading Pomeron trajectory for dipole-dipole scattering follows from a closed string exchange with $k=1$.

In~\cite{Janik:2000aj,Janik:2001sc} a result similar to (\ref{SLOPE1}) was derived for 
quark-quark scattering using a classical helicoidal surface exchange and
then corrected by one-loop bosonic quantum fluctuations. Our construction is 
physically transparent as it details the physical nature of the mechanism, and describes the produced states at the origin of the inelasticity in dipole-dipole scattering. The produced states are initially heavy extended strings of typical energy $E_L\sim b\sigma_T\sim b M_{KK}^2\lambda$ 
that ultimately decay (in $1/N_c$) to lighter closed string glueballs of energy $E_G\sim M_{KK} \lambda^0$ \cite{GLUE}. 

The Pomeron slope for dipole-dipole scattering is
$\alpha'/2$. The contribution $D_\perp/12$ in the intercept  is the L\"uscher-type contribution~\cite{Luscher}
noted in~\cite{Janik:2001sc}, although it differs by a factor of $1/8$ from our result.
Numerically, the leading Pomeron parameters of (\ref{SLOPE1}) are 
\be\label{values}
(\alpha_{{\bf P}1},\alpha_{{\bf P}1}')=(1.58,0.45\ {\rm GeV}^{-2})\,,
\ee
for $D_\perp =7$ and $\alpha'=0.9\ {\rm GeV}^{-2}$ from fit to heavy-quarkonium data. They may be compared with the values 
$(\alpha_{\bf P},\alpha_{\bf P}')=(1.08,0.25\ {\rm GeV}^{-2})$ extracted experimentally for the ``soft" Pomeron. However, our treatment assumes that the dipole size is small, so the appropriate intercept to compare with is 
the one extracted from diffractive scattering at larger values of $Q^2$ where $\alpha_{\bf P} \simeq 1.3$, see e.g. \cite{Donnachie:2001xx,Gotsman:2008tr,Ryskin:2011qe}. This is usually referred to as the "hard" Pomeron.
Also, the value $\alpha'=0.9\ {\rm GeV}^{-2}$ from heavy-quarkonium data is model-dependent, and we haven't included
the effect of flavor fermions in our analysis. Another possibility is that
the experimentally observed ``soft" Pomeron may be an effective description of multiple Pomeron exchanges as a result of unitarization, as we discuss in  section \ref{Froissart}.

\section{The Soft Pomeron}\label{DON}

The discrepancy in the intercept may be overall due to the fact that in real QCD the number of transverse fluctuations 
are effectively $D_\perp =2$ and not $D_\perp=7$ as suggested by holographic QCD. 
Indeed, L\"uscher~\cite{Luscher}
a while ago argued that long QCD strings (relevant here at large impact parameters) are described by an effective string action
of the scalar model with the least number of transverse derivatives as the dominant universal contribution. 
This scalar model is just the Polyakov action with $D_\perp=2$. As a result (\ref{DIPOLEAMPLITUDE}) simplifies to
\be
\frac 1{-2is} {{\cal T}_4(s,t)} \approx &&\frac {\pi^2 g_s^2 a^2}{2}
\sum_{k=1}^{k_{\rm max}}\frac{{(-1)^{k}}}k \,s^{1/6k+\alpha't/2k}\,
\label{4DIPOLEAMPLITUDE}
\ee
The leading Pomeron intercept is $1.16$, closer to the experimental value.

The emerging (soft) Pomeron description of our dipole-dipole analysis bears many similarities with the Donnachie-Landshoff
Pomeron~\cite{DON} for $k=1$, with 
\be
\frac 1{-2is} {{\cal T}_{\rm soft}}(s,t)\approx -(3\beta_{\bf P}\,{\bf F}_1(t))^2\,(-is\alpha_{\bf P}')^{\alpha_{\bf P}(t)-1}\,,
\label{DL}
\ee
and $\alpha_{\bf P}(t)=1.08+0.25\,t$. The dipole form-factor is normalized to ${\bf F}_1(0)=1$.  The soft Pomeron (\ref{DL})
follows from a {\it vector} coupling to the proton through $-3i\beta_{\bf P}{\bf F}_1(t)\gamma^\mu$ with a propagator 
$(-is\alpha'_{\bf P})^{\alpha_{\bf P}(t)}$ whereby $\alpha_{\bf P}$ is the Pomeron spin. From (\ref{DIPOLEAMPLITUDE})
and for $k=1$, the Pomeron-to-dipole vector coupling in holgraphy is just
\be
\beta_{\bf P}=\frac {\pi}3 g_s\,a\,\left(\frac{\pi}{{\rm ln}\,s}\right)^{D_\perp/2-1}
\ee
which is constant $\beta_{\bf P}=\pi g_s\,a/3$ for the L\"uscher scalar string model with
$D_\perp=2$. Empirically, $\beta_{\bf P}=1.87/{\rm GeV}$.

In Minkowski geometry  the dipole-dipole scattering amplitude (\ref{4X}-\ref{5X}) involves light-like Wilson lines 
sloped orthogonally on the light-cone and an important overall factor of $s$ which we have carried throughout in our analysis.
This extra factor of $s$ follows from the vector character of the gluon interaction as sourced by the light-like Wilson lines
which is at the origin of the rewriting of the S-matrix in terms of a ${\bf WW}$-correlator after LSZ reduction.  This vector
coupling is best seen in the analogue reduction of the quark-quark scattering amplitude by noting that
\be
2s&&= 2\sqrt{p_{3+}p_{1+}p_{4-}p_{2-}}\,{\bf 1}_{s_1s_3}\,{\bf 1}_{s_2s_4}\nonumber\\
&&\approx \overline{u}(s_3,p_3)\gamma^\mu u(s_1,p_1)\,\overline{u}(s_4,p_4)\gamma_\mu u(s_2,p_2)\,,
\label{2S}
\ee
for $p_{1,3\,+}, p_{2,4\,-}\rightarrow\infty$.  Also, we note that even though the $2s$ implies a C-odd coupling, the correlator
or propagator ${\bf WW}$ does not have a simple C-odd transformation so the pomeron is not a vector exchanged particle. 
Rather the ${\bf WW}$ correlator suggests that it is a world-sheet made of a coherent sum of planar gluons. In the Pomeron channel
this sum is C-even.

\section{Regime of Validity }\label{valid}

In this section, we estimate the regime of validity of our holographic approximations.
The flat space approximation where the strings are assumed to stay close to an IR end point can be tested via
the argument given in~\cite{TAN}. The string motion along the holographic direction is governed by an effective Schrodinger 
equation whose potential is somewhat model-dependent. The universal feature of this potential is that it includes a positive term proportional to $(-t)$ with a monotonic increase in the IR direction.
This implies that for a sufficiently large $(-t)$, the IR region is screened by a potential barrier, and the string worldsheets
are pushed toward the ultra-violet (UV) regime, where the description presumably goes over to a BFKL-type behavior. 
If the potential at $t=0$ has a local minimum at some point close to the IR end point, which typically happens for models with confinement and running coupling, the strings can stay close to that point for small enough $(-t)$ such that the repulsive
part proportional to $(-t)$ is subdominant. The precise analysis for our Witten's geometry is deferred to a future work, but
the prototypical analysis in~\cite{TAN} suggests that the condition for this should be

\be
\sqrt{-t} \le M_{KK}\,.
\ee
This specifies the regime of soft Pomerons that we are interested in.

Another important assumption in our calculations is neglecting worldsheet fermions.
Although the full Green-Schwarz action for models with confinement is not known yet, 
the analysis in~\cite{KINAR} showed that the Green-Schwarz worldsheet fermions get massive due to
couplings to the background Ramond-Ramond flux. In the gauge where the string worldsheet coordinates
coincide with the space-time coordinates that the string is embedded, the masses of the fermions
are shown to be roughly the mass scale of the glueballs, which is $M_{KK}$ for Witten's geometry.
We now estimate precisely when we can neglect the additional effects from these massive worldsheet fermions in our result.

The worldsheet fermions do not affect the classical worldsheet instantons and their classical action.
In the exponent of the result (\ref{XPOLES}) (it is enough to consider $k=1, n=0$ for our purpose), 
$-b^2/2\chi\alpha'$ is therefore not affected by fermions. The contribution
$D_\perp\chi/12$ to the intercept, comes from quantum 1-loop fluctuations of worldsheet bosons, and in principle worldsheet fermions can affect this result through the 1-loop determinant induced by their fluctuations

The classical instanton for $k=1$ appears at $T=2\pi/\chi$ as we saw in previous sections.
Note that the range of $\sigma$ on the worldsheet was normalized to 1 before.
In the spacetime picture, the string is extended by the distance $b$ transversely, so to match the string worldsheet coordinates
with the spacetime coordinates keeping the conformal gauge, one has to rescale the string coordinates by a factor of $b$ (Weyl transformation), so that the ranges of $(\sigma,\tau)$ are $(b,2\pi b/\chi)$ respectively. The classical Polyakov action in this gauge is
simply the area times $\sigma_T/2$ which reproduces $b^2/2\chi\alpha'$.
The mass of the worldsheet fermions is of order of $M_{KK}$ in this coordinate gauge.

The contribution to the intercept $D_\perp\chi/12$ arises as the Casimir energy or L\"uscher correction from the bosonic fluctuations on this string worldsheet. 
Note that the range of $\sigma$ is much larger than the range of $\tau$ for large $\chi$, so one should think of
$\tau$ as  space and $\sigma$ as  time in this analogy to the L\"uscher correction to the potential.
Let us consider the range of $\tau$ as an effective length $L=2\pi b/\chi$. Then $D_\perp\chi/12$ can be understood as a L\"uscher type
contribution

\be
{D_\perp \chi \over 12}={\pi D_\perp\over 6} {b\over L}=b\cdot V(L)\,.
\ee
The massive fermions of mass $M_{KK}$ contribute to the potential $V(L)$ an amount~\cite{KINAR}
\be
V(L)_F\sim D_\perp\sqrt{M_{KK}\over L}e^{-2 M_{KK}L}\,.
\ee
Using (\ref{bsaddle}), $b\sim {\rm min}\left(\sqrt{2\chi\alpha'}, {1\over\sqrt{-t}}\right)$, the condition $V_F(L)\le V(L)$ becomes equivalent to
\be
\sqrt{\chi}\ge {72\over\pi}\sqrt{27\over 2\lambda} e^{-8\pi\sqrt{27/2\lambda\chi}}\,,\label{con5}
\ee
for the case $b\sim \sqrt{2\chi\alpha'}$, and
\be
\chi\ge {72\over\pi} {M_{KK}\over\sqrt{-t}} e^{-8\pi M_{KK}/\chi\sqrt{-t}}\,,\label{con6}
\ee
for the case $b\sim 1/\sqrt{-t}$. For sufficiently large $\chi\to\infty$, both (\ref{con5}) and (\ref{con6}) are satisfied.
This justifies our approximation of neglecting the fermionic contributions to the intercept.

To compare these conditions to those discussed in~\cite{TAN},  let us restrict the analysis to 
$\sqrt{-t}=0$ for simplicity. Thus, the dominant contribution to the $b$ integration in the
scattering amplitude stems from
\be
b\sim \sqrt{2\alpha'\chi}=\sqrt{27\chi/2\lambda}M_{KK}^{-1}\,.\label{bdis}
\ee
which is the diffusion length.
This $b$ characterizes the size of a typical string worldsheet in the functional integral.
In the regime where this size is smaller than the curvature scale of the geometry at the IR end point, the strings 
feel a locally flat ten dimensional space, and the proper string theory is the critical ten-dimensional superstring theory
including worldsheet fermions. The worldsheet fermion masses become irrelevant, and 
the ``locality'' assumption in~\cite{TAN} is valid in this regime. The resulting intercept, now coming from a full critical string theory, is 
2 as in (\ref{kernel})~\cite{TAN}. This is the maximal spin of the massless states exchanged (in ten dimensions), which are the spin 2 gravitons. 

The proper length of $\Delta x^\perp\sim b$ at the IR end point in Witten's geometry is
\be
b_{\rm proper}= b\left(U_{KK}/R\right)^{3\over 4}\sim \sqrt{2\chi} l_s\,,
\ee
using (\ref{bdis}) and 
\be
R^3=\lambda l_s^2/2 M_{KK}\quad,\quad U_{KK}=2 M_{KK}\lambda l_s^2/9\,.
\ee
Here we changed the holographic coordinate in (\ref{conf_metric}) as $z=(R/U)^{3/4}$ and $z_0=(R/U_{KK})^{3/4}$. The curvature scale from the size of the internal $S^4$ is given by
\be
L_c=\left(R/U_{KK}\right)^{3\over 4} U_{KK}=\sqrt{\lambda/3} l_s\,,
\ee
so that the condition $b_{\rm proper}\le L_c$ translates to
\be
\chi\le \lambda/6\,.\label{req1}
\ee

One should also compare $b$ with the Kaluza-Klein scale $M_{KK}^{-1}$ that characterizes the 4D masses of glueballs
upon compactification to 4 dimensions. Only when $b\le M_{KK}^{-1}$, one can neglect the existence of the compact holographic direction, and justify  the ``locality'' assumption in~\cite{TAN}. From (\ref{bdis}), this yields
\be
\chi\le2\lambda/27\,,\label{req2}
\ee 
which is similar to (\ref{req1}). Only in the regime satisfying (\ref{req1}) and (\ref{req2}), 
the ``locality'' assumption and the intercept close to 2 (modulo further $1/\lambda$ corrections) in~\cite{TAN}
are justified.

As one increases $\chi$ above the bounds (\ref{req1}) and (\ref{req2}), the size of the worldsheet $b$ is large enough
to invalidate the 10 dimensional ``locality'', and the intercept starts deviating from 2. The string worldsheets start 
feeling the curvatures of the geometry, the existence of a confining scale, and the supersymmetry breaking.
Although the dynamics along the holographic direction and its effect on the intercept were studied at the border of the conditions (\ref{req1}),(\ref{req2}) (i.e. $\chi\sim\lambda$) in~\cite{TAN}, it is generally hard  to analyze these effects for $\chi\gg \lambda$. 
This limit is relevant to diffractive physics in general and the Pomeron in particular.

Qualitatively, when $b\gg M_{KK}^{-1}$ ($\chi\gg \lambda$), the masses of the 4D glueballs of order $M_{KK}$ from the Kaluza-Klein reduction of 10 dimensional gravitons are important. Indeed, the exchanged
Pomerons at low $(-t)\approx 0$ can no longer be thought of as 10 dimensional massless spin 2 
gravitons in this regime as they are replaced by 4D massive spin 2 glueballs. Whether the $t>0$ spectrum still governs the small $(-t)\ge0$ regime and thus the intercept is not a trivial question.
When (\ref{req1}) and (\ref{req2}) are satisfied, it is clear that the Regge trajectory from approximately the ``local''  10 dimensional amplitude smoothly connects $(-t)\ge 0$ and the $t>0$ spin 2 glueball spectrum as in \cite{TAN}, leading to an intercept close to 2.
However, outside the parameter range (\ref{req1}) and (\ref{req2}), it is not really possible to infer anything about $(-t)\ge0$ from the 
 $t>0$ spectrum. We do not know how to compute the full string amplitude in this regime. It is very likely that 
the intercept should come from a different physics unrelated to spin 2. Equivalently, this also means that a simple sum of supergravity field exchanges as in~\cite{Janik:1999zk} cannot give the right intercept, as these massive modes are away from the $t=0$ region.

For small $(-t)\ge 0$, we have seen that the imaginary part of the amplitude is governed by our worldsheet instantons.
What we showed was that the worldsheet theory of these instantons for large $\chi$ satisfying (\ref{con5}) and (\ref{con6}) is approximately bosonic, indicating a transition from superstring to an effective bosonic string description with a proper $D_\perp$ perhaps as advocated by L\"uscher. The resulting intercept is naturally far from 2 by a number of order 1. 
The conditions (\ref{con5}) and (\ref{con6}) requiring a large $\chi$ for the validity of this is consistent with 
the breakdown of (\ref{req1}) and (\ref{req2}) for a superstring description.
Our analysis suggests that what is responsible for a low $(-t)\approx 0 $ Pomeron trajectory 
originates from this non-critical bosonic effective string theory.

\section{Froissart Bound}\label{Froissart}

Both (\ref{DIPOLEAMPLITUDE}) and (\ref{4DIPOLEAMPLITUDE}) for the scattering amplitude
violate unitarity as $s\rightarrow\infty$. The reason is the one Pomeron exchange approximation 
to the exact dipole-dipole correlator (\ref{SCHWINGER}), which can break down for sufficiently large $s$. 
One conventional way of curing this is to sum over all multi Pomeron exchanges by exponentiating (\ref{SCHWINGER}),
neglecting inter-Pomeron interactions as an approximation.
Due to $g_s^2$ in front of (\ref{SCHWINGER}),
this is tantamount to summing a subset of $1/N_c^2$ corrections in relation to the unitarity bound.
The resulting scattering amplitude reads
\be
\frac 1{2is} {{\cal T}(s,t)} =\int\,d^2b\,e^{iq\cdot b}\,
\left(1-{\rm exp}({\bf WW}_{\rm poles})\right)\,,
\label{EXACTWW}
\ee
where the ${\bf WW}_{\rm poles}$ contribution in the exponent is given by (\ref{XPOLES}).
The total cross section follows by the optical theorem

\be
\sigma_{\rm tot}(s)=2\int\,d^2b\,
\left(1-{\rm exp}({\bf WW}_{\rm poles})\right)\,.
\label{CROSS}
\ee
Inspecting our result (\ref{XPOLES}),
the integrand in (\ref{CROSS}) becomes vanishingly small when $b\ge b_{\rm max}$ with
\be
b_{\rm max}^2={D_\perp \alpha'\over 6}\chi^2\,,
\ee
for large $s\to\infty$ limit. This is due to the fact that the contribution to ${\bf WW}_{\rm poles}$ becomes exponentially small 
when $b\gg b_{\rm max}$ from the exponential structure of (\ref{XPOLES}). On the other hand, for $b\le b_{\rm max}$,
the contribution ${\bf WW}_{\rm poles}$ becomes large and negative in the 
$s\to\infty$ limit, causing the integrand in (\ref{CROSS}) to be 1.
Therefore, $b=b_{\rm max}$ specifies a sharp transition of the integral in (\ref{CROSS}), similar to a black disc.
Thus
\be
\sigma_{\rm tot}(s)\approx 2 \int^{b_{\rm max}}\,d^2b\,= {\pi D_\perp \alpha'\over 3}\chi^2\,.
\label{BOUND}
\ee
The total cross section (\ref{BOUND}) saturates the Froissart unitarity bound, i.e. $\sigma_{\rm tot}(s)\le\chi^2$
\cite{BOUND}.

This unitarization of the scattering amplitude at high energies leads to the replacement of the ``bare" Pomeron characterized by the parameters (\ref{values}) by the ``dressed" Pomeron with a smaller effective intercept. This may be at the origin of the apparent discrepancy between the intercept (\ref{values}) and the experimental one. At very high energies, the increase of the cross section becomes logarithmic, thereby offering an alternative mechanism for the reduction of the effective intercept.

\section{Conclusions}\label{conclusion}

In holographic QCD, the inelasticity in the dipole-dipole scattering amplitude arises
from the t-channel exchange of closed bosonic strings of N-ality $k$ between the dipole world-sheets.
The Pomeron emerges as a stringy realization of the Schwinger mechanism
in the triple limit $N_c\gg \chi\gg \lambda\gg 1$ sequentially. The quantum creation process fixes not
only the Pomeron slope, but also its intercept and weight (``residue") in the elastic amplitude. Our result for
dipole-dipole scattering is similar to the one initially derived in~\cite{Janik:2000aj,Janik:2001sc} 
using semi-classical arguments for quark-quark scattering, but not identical. In particular, the Pomeron 
slope and the intercept is found to be 2 and 8 times larger respectively. The semi-classical arguments are
related to our stringy instantons of the Schwinger mechanism. In a striking way, the Schwinger pair creation process
is at the origin of the string instability observed initially in Minkowski space in~\cite{Rho:1999jm}. 

We have noted that in our analysis the Pomeron couples vectorially, and it is far from a spin 2 $t$-{\it channel} pole. 
The graviton is the starting point in the analysis of ~\cite{TAN}, which is explicit in the amplitude (\ref{kernel}) quoted in 
the introduction. The key difference between the two analyses and therefore the results is in their starting point
which is tied to the nature of the probes used. In 
our analysis which follows the initial holographic work in~\cite{Rho:1999jm,Janik:2000aj,Janik:2000pp,Janik:2001sc},
is based on the dipole-dipole scattering reduction formula (\ref{4X}-\ref{5X}) where an explicit use of the vector character
of the QCD gluon coupling was made~\cite{NACHTMANN}. The appearance of the $2s$ factor as also explained in (\ref{2S}) is a consequence of this reduction. As a result, the Pomeron exhibits a vector coupling to the dipole worldsheet from the start, which is a welcome
empirical feature of our analysis. In contrast, the closed string Virasoro-Shapiro amplitude used in~\cite{TAN} does not rely on this
reduction for the description of the Pomeron. It is an alternative way to describe the Pomeron starting from the graviton
tensorial coupling of $s^2/t$ in flat space and correcting for the spin, coupling and pole location through a $1/\lambda$
and curvature.  
Amusingly, our analysis 
breaks down at small $b$ and/or $\chi<\lambda$, while the analysis in~\cite{TAN} breaks down at large $b$ and/or
$\chi>\lambda$. 
These analyses are thus different holographic approximations to 
QCD diffraction at high energy.

In many ways our analysis is similar to that carried out by Bachas~\cite{BACHAS}
for D0-brane scattering in flat space string theory. 
While our Wilson-loop dipoles in holographic QCD are
not D-branes, the stringy bosonic exchange for the Pomeron bears some similarities.
In particular the emergence of the pole-structure in (\ref{MINFINAL}) is solely due to the 
twisted bosonic string zero modes. The effect of the twist due to rapidity is, by a T-duality transformation,
analogous to the effect of an electric field living on the dipole worldsheet, leading to a Bohm-Aharonov 
effect in the longitudinal string spectrum.

We have shown that the T-dual induced longitudinal electric field causes an Unruh acceleration and thus
an Unruh temperature. As a result, a "micro-fireball" forms which maybe at the origin of
the transverse diffusion of the string and thus of the Pomeron in the impact parameter space description. This "micro-fireball"
may explain the thermal character of multiparticle production in high energy hadronic collisions and may seed the
"firework" in AA collisions at ultrarelativistic energies as probed by RHIC and LHC. These observations will
be analysed and extended elsewhere.

We have noted that the universality arguments put forward long ago by L\"uscher~\cite{Luscher} suggest
that the use of the Polyakov action in flat space-time dimension with $D_\perp=2$ (scalar model) may be universal 
for the description of dipole-dipole scattering at large impact parameter.  The ensuing scattering amplitude bears a
leading Pomeron trajectory for $N_c=3$ and strong coupling $\lambda$ for dipoles in the fundamental representation,
which is comparable to the Pomeron trajectory inferred from experiment. We have shown that the eikonalization of the exact expression for the dipole-dipole correlator yields a total cross section for dipole-dipole scattering that is consistent with the Froissart unitarity bound.
Our results can be improved in a number of ways, for example by considering the effects of the holographic direction (curvature) in the string
propagator. Indeed, by being transverse this holographic direction is likely to exhibit transverse curved diffusion as noted
in~\cite{TAN}. This point and others will be addressed elsewhere.

\section{Acknowledgements}
We thank Gerald Dunne, Frasher Loshaj, Larry McLerran,
Edward Shuryak, Alexander Stoffers, Derek Teaney, and Kirill Tuchin for discussions.
This work was supported by the U.S. Department of Energy under Contracts No.
DE-FG-88ER40388 (GB, DK, HY and IZ) and DE-AC02-98CH10886 (GB and DK).

 \vfil

\end{document}